\documentclass[letterpaper, 10 pt, conference]{ieeeconf}  
\IEEEoverridecommandlockouts                              
\usepackage{cite}

\usepackage{amsmath}

\usepackage[margin=10mm,format=hang,singlelinecheck=false]{subfig}
\usepackage[font={small}]{caption}
\usepackage{stfloats}

\usepackage{amsfonts}

\RequirePackage[dvipsnames]{xcolor}
\definecolor{ETHcolor}{RGB}{18,105,176}
\usepackage[
  colorlinks=true,
  hyperindex=true, %
  colorlinks=true,%
  pagebackref=false,%
  plainpages=false,%
  pdfpagelabels,
  ]{hyperref}
  
\usepackage{braket}
\usepackage{makecell}
\usepackage{amsthm}
\usepackage{amssymb}

\usepackage{tikz}
\usepackage{pgfplots}
\usetikzlibrary{automata, positioning, arrows, external}
\usepgfplotslibrary{colorbrewer}
\pgfplotsset{compat=1.9}

\pgfplotsset{
	discard if not/.style 2 args={
		x filter/.code={
			\edef\tempa{\thisrow{#1}}
	            	\edef\tempb{#2}
	            	\ifx\tempa\tempb
	            	\else
	                	
	            	\fi
	        	}
	}
}

\pgfplotsset{	
	DefaultAxisStyle/.style={
        xlabel near ticks,
	    ylabel near ticks,
	    title style = {font = \small},
	    label style = {font = \small},
	    tick label style = {font = \footnotesize},
	    legend style = {font = \footnotesize,
	        legend cell align=left,
	        /tikz/every even column/.style={column sep=2pt},
	        fill=none,
	        draw=none,
	    },
	    no markers,
	    every axis plot/.append style={thick},
		table/col sep=comma,
	}
}

\pgfplotsset{	
	AlphaAxisStyle/.style={DefaultAxisStyle,
		cycle list/Set1,
		xlabel = {Future discount factor},
	}
}

\pgfplotsset{	
	CAAxisStyle/.style={DefaultAxisStyle,
		cycle list/Set1,
		xlabel = {Cost of activation},
	}
}

\pgfplotsset{	
	time_axis_style/.style={DefaultAxisStyle,
	    cycle list/Dark2,
		xlabel = {Time [days]},
	}
}

\newtheorem{theorem}{Theorem}
\newtheorem{proposition}{Proposition}
\newtheorem{definition}{Definition}

\newcommand{\Sus}{{\mathtt{S}}}
\newcommand{\Asym}{{\mathtt{A}}}
\newcommand{\Inf}{{\mathtt{I}}}
\newcommand{\Rec}{{\mathtt{R}}}
\newcommand{\Unk}{{\mathtt{U}}}
\newcommand{\St}{{\mathcal{S}}}
\newcommand{\Z}{{\mathcal{Z}}}
\newcommand{\D}{{\mathcal{D}}}
\newcommand{\A}{{\mathcal{A}}}
\newcommand{\amax}{{a_\textup{max}}}
\newcommand{\mz}{{\tilde{z}}}
\newcommand{\prob}{{\mathbb{P}}}
\newcommand{\ract}{{r_\textup{act}}}
\newcommand{\rmig}{{r_\textup{mig}}}
\newcommand{\rdis}{{r_\textup{dis}}}
\newcommand{\cmig}{{c_\textup{mig}}}
\newcommand{\cdis}{{c_\textup{dis}}}
\newcommand{\epi}{{\boldsymbol{\pi}}}
\newcommand{\sd}{{\boldsymbol{d}}}
\newcommand{\rstar}{{r_\textup{act}^*}}
\newcommand{\rbar}{{\bar{r}_\textup{act}}}
\newcommand{\pbr}{{\tilde{\pi}}}
\newcommand{\zuno}{{\mathtt{Z1}}}
\newcommand{\ztwo}{{\mathtt{Z2}}}
\newcommand{\alock}{{a_\textup{lock}}}

\newcommand{\ezz}[1]{#1}

\DeclareMathOperator*{\argmax}{argmax}

\usepackage{booktabs}

\begin{document}

\hypersetup{
  linkcolor=ETHcolor,%
  citecolor=ETHcolor,%
  filecolor=ETHcolor,%
  urlcolor=ETHcolor%
}
\hypersetup{
  pdftitle={A Dynamic Population Model of Strategic Interaction and Migration under Epidemic Risk},%
  pdfauthor={Ezzat Elokda, Saverio Bolognani and Ashish R. Hota},%
}
%
\title{\LARGE \bf A Dynamic Population Model of Strategic \\ Interaction and Migration under Epidemic Risk}

\author{Ezzat Elokda, Saverio Bolognani and Ashish R. Hota
\thanks{E. Elokda and S. Bolognani are with ETH Zurich, 8092 Zurich, Switzerland.  {\tt\small \{elokdae,bsaverio\}@ethz.ch}}
\thanks{ A. R. Hota is with the Department of Electrical Engineering, Indian Institute of Technology Kharagpur, India. {\tt\small ahota@ee.iitkgp.ac.in}}%
\thanks{The research leading to these results was partly supported by the Swiss National Science Foundation (SNSF) via the NCCR Automation.}%
}

\maketitle

\begin{abstract}
In this paper, we show how a dynamic population game can model the strategic interaction and migration decisions made by a large population of agents in response to epidemic prevalence. Specifically, we consider a modified susceptible-asymptomatic-infected-recovered (SAIR) epidemic model over multiple zones. Agents choose whether to activate (i.e., interact with others), how many other agents to interact with, and which zone to move to in a time-scale which is comparable with the epidemic evolution. We define and analyze the notion of equilibrium in this game, and investigate the transient behavior of the epidemic spread in a range of numerical case studies, providing insights on the effects of the agents' degree of future awareness, strategic migration decisions, as well as different levels of lockdown and other interventions. One of our key findings is that the strategic behavior of agents plays an important role in the progression of the epidemic and can be exploited in order to design suitable epidemic control measures.
\end{abstract}

\IEEEpeerreviewmaketitle

\section{Introduction}

Infectious diseases or epidemics spread through society by exploiting social interactions. As the disease becomes more prevalent, individuals reduce their social interaction and even migrate to safer locations in a strategic and non-myopic manner \cite{Bloomberg, Brookings}, which plays a significant role in epidemic evolution. Accordingly, past work has explored decentralized or game-theoretic protection strategies against epidemics on (static) networks \cite{trajanovski2015decentralized,hota2019game,paarporn2017networked,eksin2016disease,huang2019differential}. More recently, evolution of network topology and epidemic states in a comparable time-scale have been studied in the framework of activity-driven networks \cite{zino2017analytical,zino2020analysis}. Game-theoretic decision-making in this framework was recently studied in \cite{hota2020impacts} where myopic bounded rational agents decide whether to activate or not as a function of the epidemic prevalence.

To the best of our knowledge, there have been few rigorous game-theoretic formulations that model agents that
\begin{itemize}
\item decide their degree of activation, and consequently influence the resulting network topology,
\item decide whether to migrate to different locations, and
\item maximize both current and long-run future pay-off
\end{itemize}
in the same time-scale as epidemic evolution. In this work, we present a framework to address the above research gap.


Motivated by the presence of asymptomatic carriers in COVID-19 \cite{hu2020clinical, pang2020public}, we build upon the SAIR epidemic model studied in \cite{ansumali2020modelling,chisholm2018implications}. We consider a large population regime where the state of an individual agent is characterized by its infection state and its location (or zone). At discrete time instants, each agent decides its degree of activation and its next location. The agent is then paired randomly with other agents, and its infection state evolves following an augmented SAIR epidemic model which also takes into account \emph{unknowingly recovered} agents as described in Section \ref{sec:model}. Agents maximize a discounted infinite horizon expected reward which is a function of the aggregate infection state, the zonal distribution of the agents, and the policy followed by the population. 

In a departure from the conventional assumption of a static population distribution in the classical population game setting \cite{sandholm2010population}, epidemic evolution leads to a dynamically evolving population which makes the analysis challenging. We utilize the recently developed framework of \emph{dynamic population games}~\cite{elokda2021dynamic}, in which the authors show a reduction of the dynamic setting to a static population game setting~\cite{sandholm2010population}. This simplifies the analysis in comparison to the existing approaches such as anonymous sequential games~\cite{jovanovic1988anonymous,adlakha2015equilibria} and mean field games~\cite{huang2006large,gomes2010discrete,neumann2020stationary}, and is particularly useful for epidemic models. As a consequence of the reduction, standard evolutionary models~\cite{sandholm2010population} can be adapted for the coupled dynamics of the agents' states and strategic decision making, which evolve on the same time-scale.

The paper is structured as follows: the dynamic population model is presented in Section~\ref{sec:model}, and its stationary equilibria are analyzed in Section~\ref{sec:equilibrium}. The evolutionary update of agents' policies is modeled in Section~\ref{sec:socdynamics}. Numerical experiments reported in Section~\ref{sec:simulations} provide compelling insights into agents' behavior, effects of lockdown measures and strategic mobility patterns. For instance, we observe that if recovered agents are exempt from lockdown measures, then an increased level of activity by susceptible and asymptomatic agents can happen without having much impact on the peak and total infections. Their strategic behavior does not lead to a higher infection level and the social welfare improves due to overall higher activity levels.
\section{Model}
\label{sec:model}

We consider a homogeneous population of \ezz{non-atomic} agents or individuals. The state and the dynamics of this population are described by the following elements.

\subsection{States}

We augment the SAIR epidemic model to distinguish between recovered agents who are aware of being recovered and those who are recovered, but unaware of ever being infected. Specifically, each agent can be in one of the following infection states: \emph{Susceptible} ($\Sus$), \emph{infected Asymptomatic} ($\Asym$), \emph{Infected symptomatic} ($\Inf$), \emph{Recovered} ($\Rec$), \emph{Unknowingly recovered} ($\Unk$) (agents that have recovered without showing symptoms). Agents in states $\Rec$ and $\Unk$ are immune from further infection. Moreover, each agent resides in one of $Z$ \emph{zones} or locations. Formally, we define the state of each agent as $(s,z) \in \St \times \Z$, where $\St = \set{\Sus,\Asym,\Inf,\Rec,\Unk}$ and $\Z = \set{1,\dots,Z}$. The state distribution is $d \in \D = \Delta(\St \times \Z)$, where \ezz{$\Delta(X)$ is the space of probability distributions supported in $X$. We write} $d[s,z]$ \ezz{to denote} the proportion of agents with infection state $s$ residing in zone $z$. 

\subsection{Actions and policies}

We consider a dynamic environment that evolves in discrete-time (e.g., each time interval representing a day). At each time step, each agent strategically chooses:
\begin{itemize}
    \item its activation degree $a \in \A = \set{0,1,\dots,\amax}$, which denotes the number of other agents it chooses to interact with ($a=0$ signifies no activation), and
    \item the zone $\mz \in \Z$ where to move for the next day.
\end{itemize}
The combined action is denoted $(a,\mz) \in \A \times \Z$. A (Markovian) policy is denoted by $\pi : \St \times \Z \rightarrow \Delta(A \times \Z)$, and it maps an agent's state $(s,z) \in \St \times \Z$ to a randomization over the actions $(a,\mz) \in \A \times \Z$. 
The set of all possible policies is denoted by $\Pi$.
Explicitly, $\pi[a,\mz \mid s,z]$ is the probability that an agent plays $(a,\mz)$ when in state $(s,z)$. All agents are homogeneous and follow the same policy $\pi$. Further, agents that have never shown symptoms act in the same way, i.e.  $\pi[\cdot \mid \Sus,z] = \pi[\cdot \mid \Asym,z] = \pi[\cdot \mid \Unk,z]$. Note that $\pi$ is \emph{time-varying}; agents change their strategies as the epidemic unfolds. The dynamics of $\pi$ are detailed in Section~\ref{sec:socdynamics}.

\subsection{State transitions}
\label{ssec:statetransitions}

We now derive a dynamic model of the evolution of state distribution $d$ when the agents adopt a policy $\pi$. We denote the policy-state pair $(\pi,d)$ as the {\it social state}. The state of each agent changes at every time step according to transition probabilities encoded by the \emph{stochastic matrix}
\begin{multline}
    P[s^+,z^+ \mid s,z](\pi,d) \\= \sum_{a,\mz} \pi[a,\mz \mid s,z] \: p[s^+,z^+ \mid s,z,a,\mz](\pi,d),
    \label{eq:stochasticMatrix}
\end{multline}
where $p[s^+,z^+ \mid s,z,a,\mz](\pi,d)$ denotes the probability distribution over the next state when an agent in infection state $s$ and zone $z$ chooses action $(a,\tilde{z}$) in social state $(\pi,d)$.

Note that the Markov chain $P[s^+,z^+ \mid s,z](\pi,d)$ is not time-homogeneous as the social state $(\pi,d)$ is time-varying. The state transition function is defined as
\begin{multline}
    p[s^+,z^+ \mid s,z,a,\mz](\pi,d) \\
    := \prob[s^+ \mid s,z,a](\pi,d) \: \prob[z^+ \mid z, \mz](\pi,d). \label{eq:Transitions}
\end{multline}

The zone transition probabilities are assumed to be independent of $(\pi,d)$ and given by
\[
    \prob[z^+ \mid z, \mz](\pi,d) = \begin{cases}
        1 & \text{if } z^+ = \mz, \\
        0 & \text{otherwise}.
    \end{cases}
\]

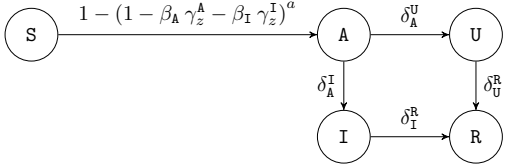
\begin{figure}[bt]
    \centering
\resizebox{0.8\columnwidth}{!}{
    \begin{tikzpicture}[
        node distance = 2.2cm,
        ->,
        >=stealth',
    ]
        \node[state] (S) {$\Sus$};
        \node[state, right of=S, xshift=3cm] (A) {$\Asym$};
        \node[state, right of=A] (U) {$\Unk$};
        \node[state, below of=A, yshift=5mm] (I) {$\Inf$};
        \node[state, right of=I] (R) {$\Rec$};
        
        \draw (S) edge[above] node{$\begin{array}{l} 1 - \left(1 - \beta_\Asym \: \gamma^\Asym_z - \beta_\Inf \: \gamma^\Inf_z\right)^a\end{array}$} (A);
        \draw (A) edge[above] node{$\delta_\Asym^\Unk$} (U);
        \draw (A) edge[left] node{$\delta_\Asym^\Inf$} (I);
        \draw (I) edge[above] node{$\delta_\Inf^\Rec$} (R);
        \draw (U) edge[right] node{$\delta_\Unk^\Rec$} (R);
    \end{tikzpicture}
    }
    \caption{Infection state transition diagram. \ezz{Self loops are not shown.}}
    \label{fig:SAIR_transitions}
\end{figure}

In order to derive the infection state transition probabilities $\prob[s^+ \mid s,z,a](\pi,d)$ (which are schematically represented in Figure~\ref{fig:SAIR_transitions}), we combine the transition rules of the (augmented) SAIR model with the specific activation actions as follows. 
\begin{itemize}
\item At a given time, an agent in state $s$ in zone $z$ chooses its activation degree $a$ according to policy $\pi$. Then, it is paired randomly with \ezz{up to} $a$ other individuals in zone $z$ with the probability of being connected with another agent being proportional to the activation degree of the target agent (analogous to the configuration model \cite{newman2010networks}).
\item \ezz{The agent could also fail to pair with one or more of the $a$ other individuals. This occurs with increasing probability as the \emph{total amount of activity} in the zone (to be defined hereafter) is low\footnote{\ezz{This represents, for example, when the public space (streets, buildings) are largely empty because most agents are staying at home.}}.} 
\item Once the network is formed, a susceptible agent becomes asymptomatic\ezz{ally infected} with probability $\beta_\Asym \in [0,1]$ per each asymptomatic neighbor and with probability $\beta_\Inf \in [0,1]$ per each infected neighbor. 
\item An asymptomatic agent becomes (symptomatically) infected\footnote{This represents both the aggravation of the illness and the outcome of testing in the asymptomatic population.} with probability $\delta_\Asym^\Inf \in (0,1]$, and recovers without being aware of it with probability $\delta_\Asym^\Unk \in [0,1]$. 
\item An infected agent recovers with probability $\delta_\Inf^\Rec \in (0,1]$.
\item An individual in state $\Unk$ becomes aware of its recovery with probability $\delta_\Unk^\Rec \in [0,1]$ (for example via serological tests on the population). 
\item The network thus formed gets discarded at the next time step and the process repeats.
\end{itemize}

Note that with the exception of the transition from state $\Sus$ to $\Asym$, all other state transition probabilities are defined via exogenous parameters and do not depend on the social state. In order to compute the transition probability from $\Sus$ to $\Asym$, we define the \emph{total amount \ezz{or mass} of activity} in zone $z$ as
\begin{equation*}
    e_z(\pi,d) = \sum_s d[s,z] \: \sum_{a,\mz} a \: \pi[a,\mz \mid s,z],
\end{equation*}
which is determined by the mass of active agents and their degrees of activation. Similarly, the mass of activity by \emph{asymptomatic} and \emph{symptomatic} agents in zone $z$ are
\begin{align*}
e^\Asym_z(\pi,d) & = d[\Asym,z] \: \sum_{a,\mz} a \: \pi[a,\mz \mid \Asym,z], \\ 
e^\Inf_z(\pi,d) & = d[\Inf,z] \: \sum_{a,\mz} a \: \pi[a,\mz \mid \Inf,z].
\end{align*}

In order to consider the event of \ezz{failing to pair with an agent} when the amount of activity in the zone $e_z(\pi,d)$ is \ezz{low}, we introduce a small constant amount $\epsilon > 0$ of fictitious activation that does not belong to any of the agents. Consequently in zone $z$, the probability of not interacting with any agent, the probability of a randomly chosen agent being asymptomatic and the probability of a randomly chosen agent being symptomatic are, respectively,
\begin{align}
\gamma^\emptyset_z(\pi,d) & = 
        \frac{\epsilon}{e_z(\pi,d) + \epsilon}, \\ 
\gamma^\Asym_z(\pi,d) & = \frac{e^\Asym_z(\pi,d)}{e_z(\pi,d) + \epsilon}, \quad  \gamma^\Inf_z(\pi,d) = \frac{e^\Inf_z(\pi,d)}{e_z(\pi,d) + \epsilon}. \label{eq:prob_gammas}
\end{align}
\ezz{Note that for a given $\epsilon$, the probability of encountering an infected agent (symptomatically or not) goes to zero as the amount of infections goes to zero, as desired.}

As a result, the probability of a susceptible agent to \emph{not} get infected upon activation \ezz{with degree $a$} is
\begin{multline*}
    \prob[s^+= \Sus \mid s = \Sus,z,a](\pi,d) \\ =\left(1 - \beta_\Asym \: \gamma^\Asym_z(\pi,d) - \beta_\Inf \: \gamma^\Inf_z(\pi,d)\right)^a.
\end{multline*}
It is easy to see that when a susceptible agent does not interact with any other agent (i.e., $a=0$), it remains susceptible. When it participates in exactly one interaction ($a=1$) in $z$, the probability that its neighbor is asymptomatic (respectively, infected) is $\gamma^\Asym_z(\pi,d)$ (respectively, $\gamma^\Inf_z(\pi,d)$). 
When it draws $a>0$ independent agents to interact with, it must not get infected in any of the interactions to remain susceptible, and this occurs with the specified probability. As a consequence, we have
\begin{align*}
& \prob[s^+= \Asym \mid s = \Sus,z,a](\pi,d) = \\
&\quad \qquad 1 - \prob[s^+= \Sus \mid s = \Sus,z,a](\pi,d).
\end{align*}
The remaining transition probabilities follow directly as: 
\allowdisplaybreaks
\begin{align*}
&\prob[s^+= \Inf \mid s = \Asym] = \delta_\Asym^\Inf, \quad \prob[s^+= \Unk \mid s = \Asym] = \delta_\Asym^\Unk, \\
& \prob[s^+= \Asym \mid s = \Asym] = 1 - \delta_\Asym^\Inf - \delta_\Asym^\Unk, \quad \prob[s^+ = \Rec \mid s = \Rec]) = 1 \\
&\prob[s^+= \Rec \mid s = \Inf] = \delta_\Inf^\Rec, \quad \prob[s^+= \Inf \mid s = \Inf] = 1 - \delta_\Inf^\Rec, \\
&\prob[s^+ = \Rec \mid s = \Unk] = \delta_\Unk^\Rec, \quad \prob[s^+ = \Unk \mid s = \Unk] = 1- \delta_\Unk^\Rec
\end{align*}
where we have suppressed $(z,a,\pi,d)$ for better readability. These expressions completely specify $\prob[s^+ \mid s,z,a](\pi,d)$ (Figure~\ref{fig:SAIR_transitions}) and therefore the state transition function~\eqref{eq:Transitions}.

\subsection{Rewards}

Each agent's own state $(s,z)$ and action $(a,\tilde z)$ yields an immediate reward for the agent, composed of a reward $\ract[s,z,a]$ for their activation decision, a reward $\rmig[s,z,\mz]$ for their migration decision, and a reward $\rdis[s]$ for how the agent's health is affected by the disease. Formally,
\begin{equation}
    r[s,z,a,\mz] := \ract[s,z,a] + \rmig[s,z,\mz] + \rdis[s]. \label{eq:Rewards}
\end{equation}

The activation reward is defined as
\[
\ract[s,z,a] := o[a] - c[s, z,a],
\]
where $o[a] \in \mathbb{R}_{+}$ denotes the social benefit of interacting with $a$ other agents and is assumed to be non-decreasing in $a$ with $o[0]=0$, and $c[s,z,a] \in \mathbb{R}_{+}$ denotes the cost imposed by the authority in zone $z$ to discourage activity.
We assume that $c[s,z,a]$ are non-decreasing in $a$ and satisfy
    \[
    \!\!\!c[\Inf,z,a] \ge c[\Sus,z,a] = c[\Asym,z,a] = c[\Unk,z,a] \ge c[\Rec,z,a]
    \]
element-wise, since lockdown measures can be more stringent against individuals showing symptoms and more benign for individuals who are known to be immune.

The migration reward encodes the non-negative cost of migrating to a new zone. In this work, we define
\[
    \rmig[s,z,\mz] := \begin{cases}
        0 & \text{if } \mz = z, \\
        -\cmig & \text{if } \mz \neq z.
    \end{cases}
\]
However, one may consider a richer cost function that incorporates specific travel restrictions between zones, etc. The third term in \eqref{eq:Rewards} encodes the cost of being ill:
\[
\rdis[s] := \begin{cases}
- \cdis & \text{if }s = \Inf, \\
0 & \text{otherwise}.
\end{cases}
\]

\subsection{Agents' Strategic Decisions}

We now introduce the strategic decision-making process of the agents. The immediate expected reward of an agent in state $(s,z)$ when it follows policy $\pi$ is
\begin{equation*}
        R[s,z](\pi) = \sum_{a,\mz} \pi[a,\mz \mid s,z] \: r[s,z,a,\mz],
\end{equation*}
with $r[s,z,a,\mz]$ as defined in~\eqref{eq:Rewards}. The expected discounted infinite horizon reward of an agent in state $(s,z)$ with discount factor $\alpha \in [0,1)$ following the homogeneous policy $\pi$ is recursively defined as
\begin{multline*}
    V[s,z](\pi,d) = R[s,z](\pi) \\+ \alpha \sum_{s^+,z^+} P[s^+,z^+ \mid s,z](\pi,d) \: V[s^+,z^+](\pi,d), 
\end{multline*}
or, equivalently in vector form,
\begin{equation}
    V(\pi,d) = (I - \alpha \: P(\pi,d))^{-1} \: R(\pi). \label{eq:V-function}
\end{equation}
\ezz{Equation~\eqref{eq:V-function} is the well-known Bellman equation.} Note that \ezz{it} is continuous in the social state $(\pi,d)$, as $I - \alpha \: P(\pi,d)$ is guaranteed to be invertible for $\alpha \in [0,1)$.

While an agent can compute the expected discounted reward $V(\pi,d)$ at a given social state $(\pi,d)$, the policy $\pi$ may not be optimal for the agent. Thus, we assume that each agent chooses its current action $(a,\mz)$ in order to maximize
\begin{multline}
        Q[s,z,a,\mz](\pi,d) := r[s,z,a,\mz] \\+ \alpha \sum_{s^+,z^+} p[s^+,z^+ \mid s,z,a,\mz](\pi,d) \: V[s^+,z^+](\pi,d),
        \label{eq:definitionQ}
\end{multline}
i.e., the agent is aware of the immediate reward and the effect of its action on their future state; however, it assesses the future reward based on a stationarity assumption on the social state $(\pi,d)$. In other words, the agent chooses its action to maximize a single-stage deviation from the homogeneous policy $\pi$~\cite[Section 2.7]{filar2012competitive}, and assumes that its own actions are not going to affect the social state significantly. 
\section{Equilibrium Analysis}
\label{sec:equilibrium}

We start by introducing the notion of best response based on the single-stage deviation reward defined in \eqref{eq:definitionQ}.

\begin{definition}[Best Response] The best response of an agent in state $(s,z)$ at the social state $(\pi,d)$ is the set valued correspondence $B_{s,z} : \Pi \times \D \rightrightarrows \Delta(\A \times \Z)$ given by
\begin{multline}
    \label{eq:BestResponse}
    B_{s,z}(\pi,d) \in \Big\{\sigma \in \Delta(\A \times \Z) : \forall \sigma' \in \Delta(\A \times \Z) \\ \sum_{a,\mz} \left(\sigma[a,\mz] - \sigma'[a,\mz] \right)\: Q[s,z,a,\mz](\pi,d) \geq 0 \Big\}.
\end{multline}
\end{definition}

The above notion of best response is from the perspective of an individual agent in state $(s,z)$ when all other agents are following the homogeneous policy $\pi$ and their states are distributed as per $d$. The agent will choose any distribution $\sigma$ over the actions which maximizes its expected single-stage deviation reward $Q$ at the current state $(s,z)$. 

Consequently, a social state $(\epi,\sd)$ is stationary when agents in all states are playing their best response when they follow the policy $\epi$, and  the state distribution $\sd$ is stationary under this policy. Thus, we have the following definition of a stationary equilibrium. 

\begin{definition}[Stationary equilibrium]
A stationary equilibrium is a social state $(\epi,\sd) \in \Pi \times \D$ which satisfies
\begin{align}
    \epi[\cdot \mid s,z] &\in B_{s,z}(\epi,\sd), \; \forall (s,z) \in \St \times \Z, \label{eq:SE-1} \tag{SE.1}\\
    \sd &= \sd \: P(\epi,\sd). \label{eq:SE-2} \tag{SE.2}
\end{align}
\end{definition}

Thus, at the equilibrium, the state Markov chain $P(\epi,\sd)$ \eqref{eq:stochasticMatrix} is time-homogeneous, and the agents behave optimally in the corresponding Markov decision process~\cite{filar2012competitive}. \ezz{Note that we denote stationary equilibria with boldface notation.} \ezz{When $(\epi,\sd)$ is a stationary equilibrium, $\epi$ corresponds to the Nash equilibrium under state distribution $\sd$.} 

\begin{theorem}[Theorem~1 in \cite{elokda2021dynamic}]
A stationary equilibrium $(\epi,\sd)$ for the proposed dynamic population game is guaranteed to exist.
\end{theorem}

We refer to~\cite{elokda2021dynamic} for the details of the proof, which relies on a fixed-point argument. \ezz{There, a general dynamic population game is considered in which the state and action spaces are finite, and} the state transition and reward functions are continuous in the social state $(\pi,d)$. \ezz{By definition, our model is an instance of such a dynamic population game.}

The following proposition shows that the stationary distribution $\sd$ does not have any asymptomatic or infected agents (i.e., eventually, the epidemic dies out). The final stationary distribution is however not unique.

\begin{proposition}
\label{pro:characterizationeq}
Let ${r}_\textup{act}^*[s,z] := \max_a \ract[s,z,a]$ be the maximum activation reward in state $s$ and zone $z$, and let ${\mathcal A}_{s,z}^*$ be the set of activation levels that achieve such reward.
Let $\bar{r}_\textup{act}[s] := \max_z {r}_\textup{act}^*[s,z]$ be the maximum activation reward in state $s$ across all zones, and let $\delta_\Unk^\Rec, \alpha, \cmig > 0$. 
Let us define the subset of zones
\begin{align*}
\bar {\mathcal Z}_s & = \argmax_z {r}_\textup{act}^*[s,z], 
\\ 
    \mathcal Z^0_s & = \left\{ z \mid
     \bar{r}_\textup{act}[s] - {r}_\textup{act}^*[s,z] > \frac{1-\alpha}{\alpha} \cmig   
    \right\}.
\end{align*}
Then, any social state $(\epi,\sd)$ satisfying
\begin{equation*}
    \begin{cases}
        \sd[\Asym,z] = \sd[\Inf,z] = \sd[\Unk,z] = 0 & \forall z, \\
        \sd[s,z \in \mathcal Z^0_s] = 0 &\forall s, \\
        \epi[a\notin \mathcal A^*_{s,z}, \mz \mid s,z] = 0 & \forall \mz, s, z, \\
        \epi[a, \mz\notin \bar{\mathcal Z}_s \mid s,z \in \mathcal Z^0_s] = 0 & \forall a, s,\\
        \epi[a, \mz \neq z \mid s,z \notin \mathcal Z^0_s] = 0 & \forall a, s.
    \end{cases}
\end{equation*}
is a stationary equilibrium.
\end{proposition}

The proof is presented in the extended version \cite{elokdadynamic_cdc_arxiv}. Proposition~\ref{pro:characterizationeq} shows how stationary equilibria can be computed without solving a fixed-point problem, and directly identifies some dominant strategies for the agents.
The identification of dominant strategies is insightful for the design of interventions (e.g., lockdown measure for the different compartments).
For example, it is possible to verify that
\[
\mathcal A_{\Rec,z}^* = 
\argmax_a\ 
\ract[\Rec, z, a] = 
\argmax_a\ 
\left(o[a] - c[\Rec, z, a]\right)
\]
corresponds to the dominant activation strategies for agents in $\Rec$ (\emph{knowingly} immune agents).
Notice that the activation caused by immune agents appears in the denominator of the probability that a generic agent interacts with an infectious agent -- see \eqref{eq:prob_gammas} -- \ezz{and} therefore looser lockdown measures for recovered agents can be used to reduce the spreading as shown in simulations in Section~\ref{sec:simulations}.
\section{Social State Dynamics} 
\label{sec:socdynamics}

Different stationary equilibria correspond to drastically different outcomes in terms of the impact of the epidemic on the population.
For this reason, we investigate the \emph{transient behavior} leading to the equilibrium, which is determined by the state dynamics from Section~\ref{ssec:statetransitions}, i.e., $d^+ = d \: P(\pi,d)$, and the way in which agents update their policies. For this second part, we get inspiration from the \emph{evolutionary dynamic} models in classical population games~\cite{sandholm2010population}, and more precisely from the \emph{perturbed best response dynamics}.

We assume that the agents are not perfectly rational, with the bounded rationality factor $\lambda \in [0,\infty)$. When they are making a decision on which action to play, they follow the \emph{logit choice} function~\cite[Section~6.2]{sandholm2010population}, given by
\[
    \pbr[a,\mz \mid s,z](\pi,d) = \frac{\exp{(\lambda \: Q[s,z,a,\mz](\pi,d))}}{\sum_{a',\mz'} \exp{(\lambda \: Q[s,z,a',\mz'](\pi,d))}}.
\]
For $\lambda = 0$, it results in a uniform distribution over all the actions. At the limit $\lambda \rightarrow \infty$, we recover the perfect best response. At finite value of $\lambda$, it assigns higher probabilities to actions with higher payoffs. 

In order to model the fact that agents update their policies gradually, we consider the discrete-time update 
\[
    \pi^+[\cdot \mid s,z] = (1 - \eta) \: \pi[\cdot \mid s,z] + \eta \: \pbr[\cdot \mid s,z],
\]
where $\eta \in (0,1]$ is a parameter that controls the rate of policy change: for $\eta < 1$, agents have \emph{inertia} in their decision making, while for $\eta = 1$ agents promptly update their action decision to the perturbed best response $\pbr$.
Note that this update model leads to a perturbed version of the equilibrium policy $\epi$ at the rest points, rather than the exact policy~\cite{sandholm2010population}.
\section{Numerical Case Studies}
\label{sec:simulations}

We present a select number of case studies to showcase 
\begin{itemize}
    \item the effect of agents' strategic activation decisions on the spread of the epidemic, 
    \item the impact of lockdown measures on both epidemic containment and social welfare, and
    \item the effect of strategic migration decisions on how the epidemic spreads across multiple locations.
\end{itemize}

For this purpose, we consider an infectious epidemic characterized by $\beta_\Asym = \beta_\Inf = 0.2$, $\delta_\Asym^\Inf = \delta_\Asym^\Unk = 0.08$, and $\delta_\Inf^\Rec=0.04$.
The agents can activate up to degree $\amax=6$, and the activation reward is linear in the activation degree, with a unit reward for maximum activation $o[\amax] = 1$.
The illness is quite severe, with a discomfort cost $\cdis = 10$. Initially, we let the agents choose an activation degree uniformly at random and do not plan any move. The agents are highly rational ($\lambda=10$) and unless otherwise stated, we consider that they update their decisions with an inertia $\eta=0.2$. We consider both a single zone and a two zone setting, and denote the zones by $\zuno$ and $\ztwo$.
In all cases, the epidemic starts in $\zuno$ with a proportion of $2\%$ of that zone's population asymptomatic ($\Asym$), and $1\%$ infected ($\Inf$).

We further consider that authorities can enforce lockdown regulations through the parameter $\alock[s,z]$, which represents the \emph{maximum allowed activation degree} and can differ between zones and for agents in different infection states.
Lockdown is implemented by setting $c[s,z,a] = 0$ if $a \leq \alock[s,z]$, and $c[s,z,a] = 3\:o[a]$ otherwise.
Regardless of the lockdown measures, we always assume that the discomfort of the illness is sufficient to prevent symptomatically infected agents from activating. As a consequence, the main threat of the epidemic is due to the presence of asymptomatically infected agents in the population.

\subsection{Strategic activation under lockdown measures}

\begin{figure*}[t]
    \raggedleft
    \subfloat{
        \includegraphics[trim=4mm 0 0 0 ]{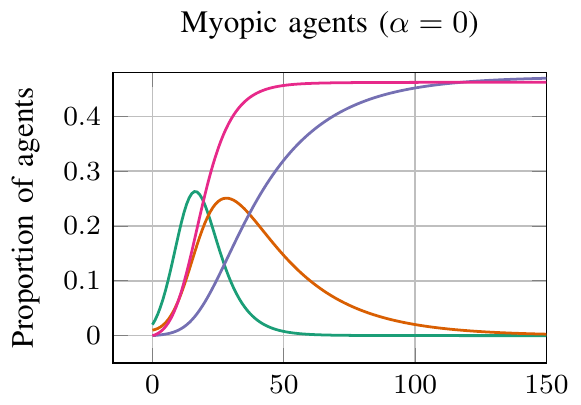}
        \label{fig:a_lock2-alpha0-delta_U_R0-R-lock-d}
    }
    \hfil
    \subfloat{
        \includegraphics{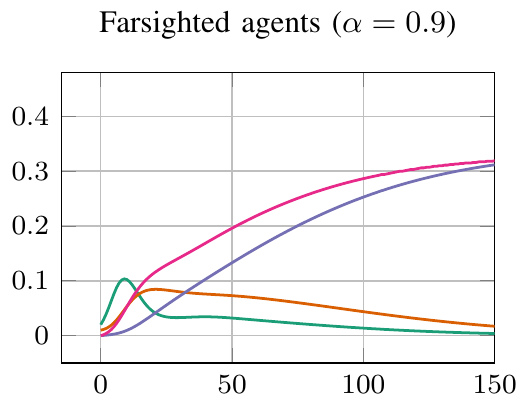}
        \label{fig:a_lock2-alpha0.9-delta_U_R0-R-lock-d}
    }
    \hfil
    \subfloat{
        \includegraphics{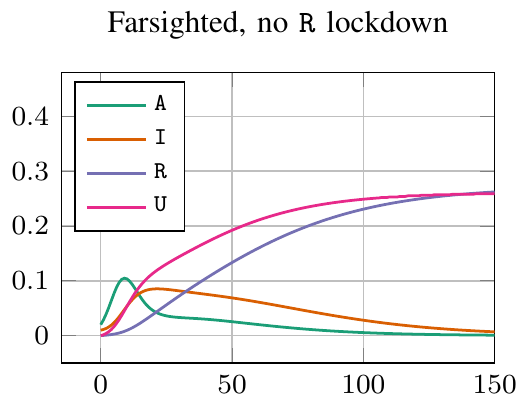}
        \label{fig:a_lock2-alpha0.9-delta_U_R0-d}
    }
    
    \setcounter{subfigure}{0}
    \subfloat[][Total infections: $93.4\%$ \\ Peak infections: $25.1\%$]{
        \includegraphics[trim=4mm 0 0 0 ]{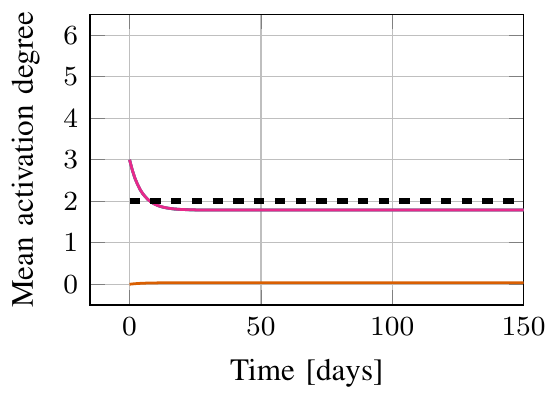}
        \label{fig:a_lock2-alpha0-delta_U_R0-R-lock}
    }
    \hfil
    \subfloat[][Total infections: $67.0\%$ \\ Peak infections: $8.4\%$]{
        \includegraphics{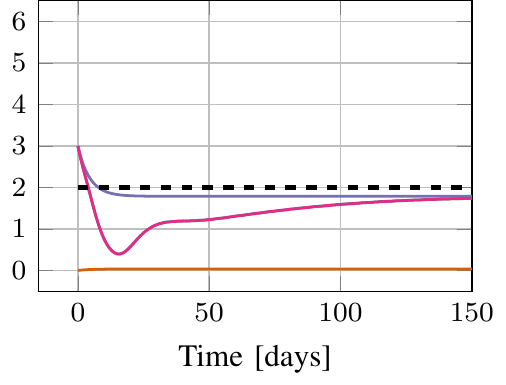}
        \label{fig:a_lock2-alpha0.9-delta_U_R0-R-lock}
    }
    \hfil
    \subfloat[][Total infections: $53.2\%$ \\ Peak infections: $8.5\%$]{
        \includegraphics{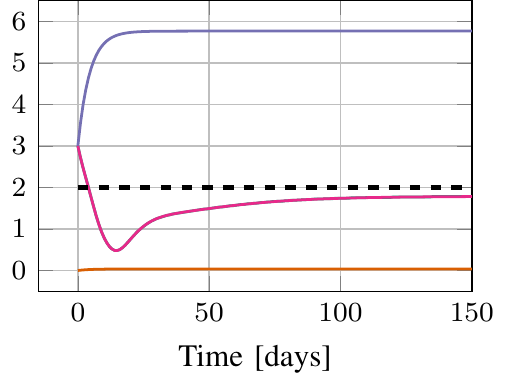}
        \label{fig:a_lock2-alpha0.9-delta_U_R0}
    }
    \caption{The effect of the agents' farsightedness and of the exclusion of immune agents from lockdown measures on the epidemic. \ezz{The dashed black lines indicate the maximum allowed activation degree.}}
    \label{fig:lockdown-trends}
\end{figure*}

We first investigate the single zone scenario, with a focus on how agents of different cognitive ability react under various lockdown measures, and the resulting effects on the epidemic spread.
Figure~\ref{fig:lockdown-trends} shows an example with lockdown degree $\alock=2$ and three cases.
In cases~\eqref{fig:a_lock2-alpha0-delta_U_R0-R-lock} and~\eqref{fig:a_lock2-alpha0.9-delta_U_R0-R-lock}, the lockdown is enforced on the whole population, and the cases differ in the agents' cognitive level of the future.
In~\eqref{fig:a_lock2-alpha0-delta_U_R0-R-lock}, the agents are completely myopic.
Notice how they simply adhere to the lockdown degree\footnote{\label{ft:pbr}A slight deviation to the expected degree of activation is due to the agents' bounded rationality.}, which aligns with standard epidemic models.
Farsighted agents~\eqref{fig:a_lock2-alpha0.9-delta_U_R0-R-lock}, on the other hand, actively adjust their activation decisions in response to the epidemic threat, and volunteer to limit their activity beyond the lockdown requirement at peak infection times. The reduction in activity levels leads to a less severe epidemic spread, with a smaller total and peak infection.

\begin{figure*}[bt]
    \centering
    \subfloat{
        \includegraphics{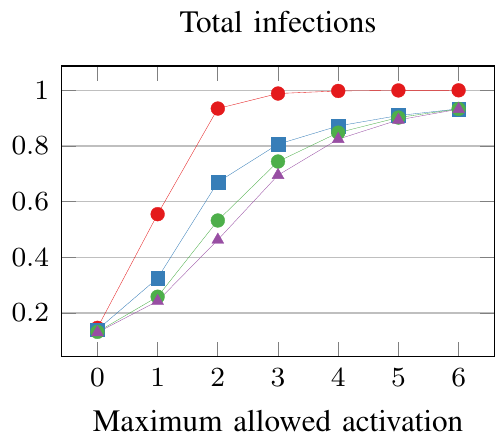}
        \label{fig:lockdown-d_RU}
    }
    \hfil
    \subfloat{
        \includegraphics{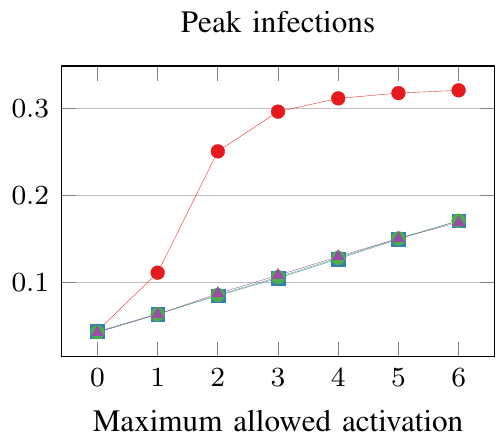}
        \label{fig:lockdown-max_d_I}
    }
    \hfil
    \subfloat{
        \includegraphics{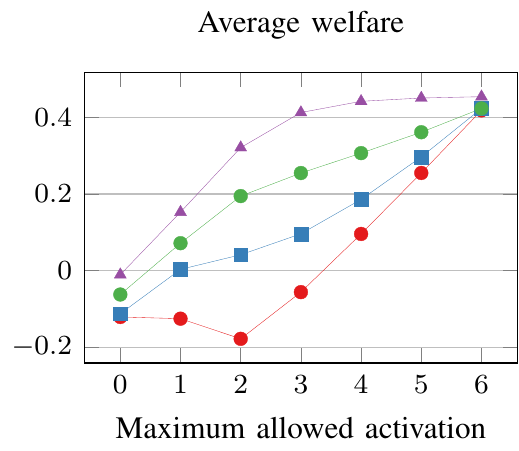}
        \label{fig:lockdown-mean_R}
    }
    
    \vspace*{-3mm}
    \subfloat{
        \includegraphics{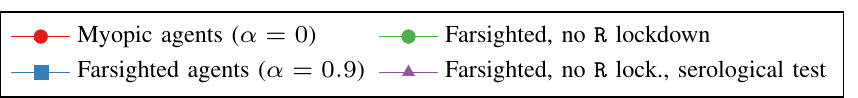}
        \label{fig:lockdown-legend}
    }
    \caption{The effect of the severity of the lockdown measures on three key epidemic indicators, for different scenarios.}
    \label{fig:lockdown-performance}
\end{figure*}

Case~\eqref{fig:a_lock2-alpha0.9-delta_U_R0} also considers farsighted agents, but this time \emph{recovered} agents are exempt from lockdown and thus activate at the maximum degree, as per their dominant strategy.
This leads to a significant reduction of the total amount of infections. In fact, due the prevalence of activity of immune agents, it becomes less likely for a susceptible agent to encounter an infected agent. Consequently, susceptible, asymptomatic and unknowingly recovered agents too increase their level of activation. Nevertheless, the peak infection remains being largely unchanged.

This insight is further explored in Figure~\ref{fig:lockdown-performance}, which shows the effect of different lockdown measures on three main performance indices:
the \emph{total infections} ($\Rec+\Unk$ at the end of the epidemic), 
the \emph{peak infections} (highest value of $\Inf$\footnote{We do not consider infected but asymptomatic agents since they do not contribute to the load on medical facilities.}),
and the \emph{average welfare} (mean reward~\eqref{eq:Rewards} in the population over the duration of the epidemic).

We depict the same cases considered in Figure~\ref{fig:lockdown-trends} and perform a parameter sweep over the strictness of the lockdown $\alock$.
Additionally, we showcase the effect of performing \emph{serological tests} to increase the amount of knowingly immune agents, at rate $\delta_\Unk^\Rec=0.05$.
We observe that myopic agents perform poorly along all the performance metrics. 
For farsighted agents, the exemption of recovered agents and serological tests lead to significant improvements in the average welfare because the knowingly immune agents are able to achieve their maximum activation degree without increasing the threat of the epidemic.

\subsection{Strategic migration}

We showcase the effect of strategic migration in a setting with two zones, with zone $\zuno$ initially holding $90\%$ of the total population, and zone $\ztwo$ initially infection free.
In both zones, (knowingly) recovered agents are exempt from lockdown, and the lockdown restrictions for the other agents are different in the zones. Namely, $\zuno$ has a looser lockdown, with a maximum allowed activation degree of 4, whereas $\ztwo$ only allows a maximum of 2. The migration cost is $\cmig=2$, and the agents are farsighted with $\alpha=0.9$.
Note that with these parameters, susceptible agents will want to move to $\zuno$ when the epidemic is not prevalent, as per Proposition~\ref{pro:characterizationeq}\footnote{Here, $\rstar[\Sus,\ztwo]=\frac{1}{3}$ and $\rbar[\Sus]=\frac{2}{3}$, therefore $\ztwo \in \Z^0_\Sus$.}.
The inertia in the policy update is $\eta=0.1$.

\begin{figure*}[t]
    \centering
    \subfloat{
        \includegraphics{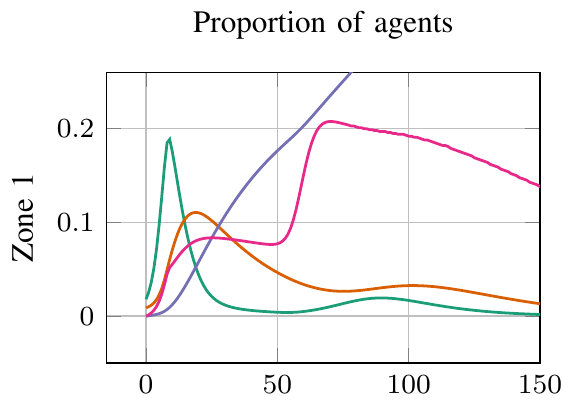}
        \label{fig:migration-zone-1-d}
    }
    \hfil
    \subfloat{
        \includegraphics{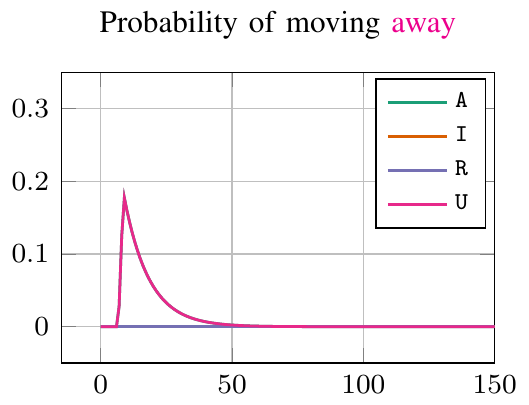}
        \label{fig:migration-zone-1-p_m}
    }
    \hfil
    \subfloat{
        \includegraphics{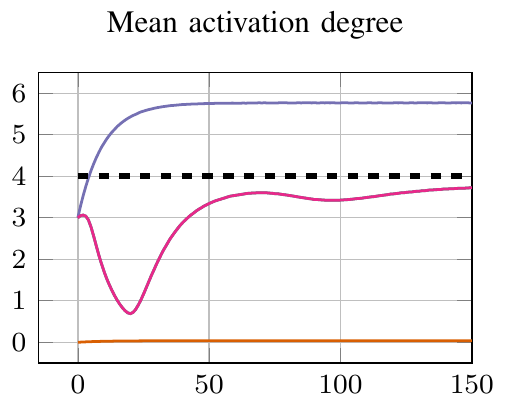}
        \label{fig:migration-zone-1-mean_a}
    }
    
    \subfloat{
        \includegraphics{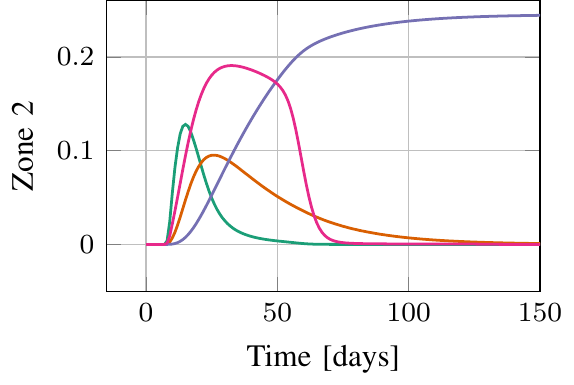}
        \label{fig:migration-zone-2-d}
    }
    \hfil
    \subfloat{
        \includegraphics{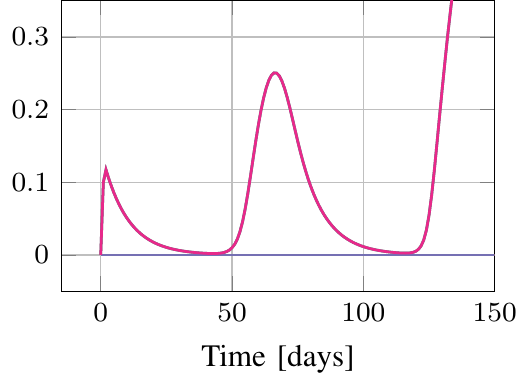}
        \label{fig:migration-zone-2-p_m}
    }
    \hfil
    \subfloat{
        \includegraphics{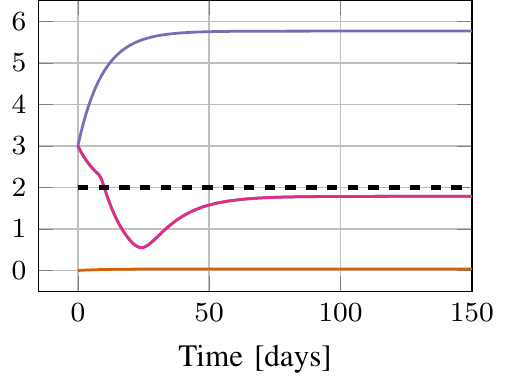}
        \label{fig:migration-zone-2-mean_a}
    }
    \caption{Effect of strategic migration/activation decisions on the epidemic spreading between two zones with different lockdown measures. \ezz{The dashed black lines indicate the maximum allowed activation degree. The top left and bottom center plots are clipped for display purposes; at the end of the epidemic all non-recovered agents in Zone 2 move to Zone 1. The total infections in Zones 1 and 2 are, respectively, 62.1\% and 24.5\%. The peak infections in Zones 1 and 2 are, respectively, 11.1\% and 9.5\%.}}
    \label{fig:migration-trends}
\end{figure*}

Additionally, both zones perform serological testing at rate $\delta_\Unk^\Rec=0.01$.
The resulting epidemic spread is displayed in Figure~\ref{fig:migration-trends}.
First, notice how the proportion of unknowingly recovered agents decays, in contrast to Figure~\ref{fig:lockdown-trends}, in which no serological testing is performed.

We now focus on the strategic migration behavior of agents who are either susceptible or think they are ($\Sus$, $\Asym$, $\Unk$). Note that symptomatically infected and recovered agents never move since they are immune to the threat of the epidemic, and the activation costs are the same for them in both zones.
Initially, the epidemic risk is still small in $\zuno$, and the occupants of $\ztwo$ start moving there to benefit from the more lenient lockdown measure.
This trend soon reverses, however, with the rise of infections in $\zuno$: the strategic agents elect to move to the zone with stricter lockdown to escape the epidemic risk.
Since a proportion of the movers are asymptomatically infected, this leads to an outbreak of the epidemic in $\ztwo$ as well, with lower, but significant, peak infections than $\zuno$.
Eventually, once the infections in $\zuno$ has decreased sufficiently (at approximately day 50), some $\ztwo$ residents move to $\zuno$ again, initiating a \emph{second wave of infections} in $\zuno$. At the end of the epidemic, all the remaining agents move from $\ztwo$ as per their dominant strategy. 

\section{Conclusions and Future Works}

In this paper, we propose a model of strategic behavior at both individual and societal level based on first principles, demonstrate its potential to explain complex activation and migration patterns, and to guide the design of more effective epidemic control measures. We characterize the stationary equilibria in the proposed dynamic population game setting and illustrate how a better understanding of the emergent behavior can be leveraged to design effective mitigation strategies. For instance, we show that withdrawing restrictions on recovered agents leads to higher levels of activity by susceptible agents without increasing the peak and total infection levels. This observation provides a rigorous justification for conducting large-scale serological testing and letting individuals with immunity to interact freely for significantly improving the welfare of the society. A natural extension of this model includes considering the behavior of vaccinated individuals and designing optimal intervention strategies that incorporates the strategic response of agents. 
\appendix

\subsection*{Proof of Proposition~\ref{pro:characterizationeq}}

The first property follows the fact that states $\Asym$, $\Inf$, and  $\Unk$  are \emph{transient} in the infection state Markov chain (see Figure~\ref{fig:SAIR_transitions}), and evolve independently of the agents' migrations. We omit the formal steps, \ezz{which follow standard Markov chain theory arguments}.

We now observe that, given a distribution that is only supported on the two compartments $\Sus$ and $\Rec$, no transition between compartments of the extended SAIR model is possible. The reward for agents in both these compartments then becomes independent from the actions of others, and the set $\mathcal A^*_{s,z}$ defines their \emph{dominant activation strategies}. 

We now consider migration strategies of the agents at the equilibrium $(\epi,\sd)$. We first consider an agent in infection state $s \in \{\Sus,\Rec\}$ residing in zone $z \in \bar {\mathcal Z}_s$. In this zone, the activation reward is maximum among all other zones for the agent in state $s$. Since the infection state remains unchanged, and migration is costly ($\cmig>0$), migrating to a different zone does not lead to a beneficial single-stage deviation for the agent.

We now consider an agent in zone $z \in {\mathcal Z}^{n}_s := {\mathcal Z} \setminus \{\bar {\mathcal Z}_s \cup \mathcal Z^0_s\}$. According to policy $\epi$, the agent does not migrate to a different zone. Consequently, the value function for the agent is
\[
    V[s,z](\epi,\sd) = \frac{{r}_\text{act}^*[s,z]}{1-\alpha}.
\]
Now, consider a single-stage deviation where the agent moves to location $z' \in \bar {\mathcal Z}_s$ and chooses activation degree $a' \in \mathcal{A}^*_{s,z}$. Consequently, using the definition of ${\mathcal Z}^{n}_s$,
\[
 Q[s,z,a',z'] =  r^*_\text{act}[s,z]-\cmig + \alpha \frac{\bar{r}_\text{act}[s]}{1-\alpha} < V[s,z](\epi,\sd).
\]
In other words, an agent in a zone in ${\mathcal Z}^{n}_s$ does not find it beneficial to migrate anywhere else in a single-stage deviation. 

It remains to show that an agent in zone $z \in \mathcal{Z}^0_s$ finds it beneficial to move to a zone in $\bar{\mathcal{Z}_s}$ and consequently, we must have $d[s,z]=0$ for $z \in \mathcal{Z}^0_s$. Under policy $\epi$, we have
$$ \epi[a \notin \mathcal{A}^*_{s,z}, z' \notin \bar{\mathcal{Z}}_s|s,z] = 0.$$
Consequently, the value function for the agent is
\begin{align*}
V[s,z](\epi,\sd) & = r^*_\text{act}[s,z] - \cmig + \alpha \frac{\bar{r}_\text{act}[s]}{1-\alpha} 
> \frac{r^*_\text{act}[s,z]}{1-\alpha};
\end{align*}
in other words, the policy $\epi$ yields a higher value compared to any policy that does not include migration. Similarly, one can show that migrating to a zone in ${\mathcal Z}^{n}_s$ instead does not yield a beneficial single-stage deviation since $r^*_\text{act}[s,z'] < \bar{r}_\text{act}[s]$ for any zone $z' \in {\mathcal Z}^{n}_s$. \qed

\bibliographystyle{IEEEtran}
\bibliography{IEEEabrv,bibliography.bib,refs_new.bib}

\begin{thebibliography}{10}
\providecommand{\url}[1]{#1}
\csname url@samestyle\endcsname
\providecommand{\newblock}{\relax}
\providecommand{\bibinfo}[2]{#2}
\providecommand{\BIBentrySTDinterwordspacing}{\spaceskip=0pt\relax}
\providecommand{\BIBentryALTinterwordstretchfactor}{4}
\providecommand{\BIBentryALTinterwordspacing}{\spaceskip=\fontdimen2\font plus
\BIBentryALTinterwordstretchfactor\fontdimen3\font minus
  \fontdimen4\font\relax}
\providecommand{\BIBforeignlanguage}[2]{{%
\expandafter\ifx\csname l@#1\endcsname\relax
\typeout{** WARNING: IEEEtran.bst: No hyphenation pattern has been}%
\typeout{** loaded for the language `#1'. Using the pattern for}%
\typeout{** the default language instead.}%
\else
\language=\csname l@#1\endcsname
\fi
#2}}
\providecommand{\BIBdecl}{\relax}
\BIBdecl

\bibitem{Bloomberg}
\BIBentryALTinterwordspacing
M.~Patino, A.~Kessler, and S.~Holder. (2021) More {A}mericans are leaving
  cities, but don’t call it an urban exodus. Accessed: 2021-08-26. [Online].
  Available:
  \url{https://www.bloomberg.com/graphics/2021-citylab-how-americans-moved/}
\BIBentrySTDinterwordspacing

\bibitem{Brookings}
\BIBentryALTinterwordspacing
W.~H. Frey. (2021) Pandemic population change across metro {A}merica:
  {A}ccelerated migration, less immigration, fewer births and more deaths.
  Accessed: 2021-08-26. [Online]. Available: \url{https://brook.gs/3hFPAFf}
\BIBentrySTDinterwordspacing

\bibitem{trajanovski2015decentralized}
S.~Trajanovski, Y.~Hayel, E.~Altman, H.~Wang, and P.~Van~Mieghem,
  ``Decentralized protection strategies against {SIS} epidemics in networks,''
  \emph{IEEE Transactions on Control of Network Systems}, vol.~2, no.~4, pp.
  406--419, 2015.

\bibitem{hota2019game}
A.~R. Hota and S.~Sundaram, ``Game-theoretic vaccination against networked
  {SIS} epidemics and impacts of human decision-making,'' \emph{IEEE
  Transactions on Control of Network Systems}, vol.~6, no.~4, pp. 1461--1472,
  2019.

\bibitem{paarporn2017networked}
K.~Paarporn, C.~Eksin, J.~S. Weitz, and J.~S. Shamma, ``Networked {SIS}
  epidemics with awareness,'' \emph{IEEE Transactions on Computational Social
  Systems}, vol.~4, no.~3, pp. 93--103, 2017.

\bibitem{eksin2016disease}
C.~Eksin, J.~S. Shamma, and J.~S. Weitz, ``Disease dynamics on a network game:
  {A} little empathy goes a long way,'' \emph{Scientific Reports}, vol.~7, p.
  44122, 2017.

\bibitem{huang2019differential}
Y.~Huang and Q.~Zhu, ``A differential game approach to decentralized
  virus-resistant weight adaptation policy over complex networks,'' \emph{IEEE
  Transactions on Control of Network Systems}, vol.~7, no.~2, pp. 944--955,
  2020.

\bibitem{zino2017analytical}
L.~Zino, A.~Rizzo, and M.~Porfiri, ``An analytical framework for the study of
  epidemic models on activity driven networks,'' \emph{Journal of Complex
  Networks}, vol.~5, no.~6, pp. 924--952, 2017.

\bibitem{zino2020analysis}
------, ``Analysis and control of epidemics in temporal networks with
  self-excitement and behavioral changes,'' \emph{European Journal of Control},
  vol.~54, pp. 1 -- 11, 2020.

\bibitem{hota2020impacts}
A.~R. Hota, T.~Sneh, and K.~Gupta, ``Impacts of game-theoretic activation on
  epidemic spread over dynamical networks,'' \emph{arXiv:2011.00445}, 2020.

\bibitem{hu2020clinical}
Z.~Hu, C.~Song, C.~Xu, G.~Jin, Y.~Chen, X.~Xu, H.~Ma, W.~Chen, Y.~Lin, Y.~Zheng
  \emph{et~al.}, ``Clinical characteristics of 24 asymptomatic infections with
  {COVID}-19 screened among close contacts in {N}anjing, {C}hina,''
  \emph{Science China Life Sciences}, vol.~63, no.~5, pp. 706--711, 2020.

\bibitem{pang2020public}
W.~Pang, ``Public health policy: {COVID-19} epidemic and {SEIR} model with
  asymptomatic viral carriers,'' \emph{arXiv:2004.06311}, 2020.

\bibitem{ansumali2020modelling}
S.~Ansumali, S.~Kaushal, A.~Kumar, M.~K. Prakash, and M.~Vidyasagar,
  ``Modelling a pandemic with asymptomatic patients, impact of lockdown and
  herd immunity, with applications to sars-cov-2,'' \emph{Annual Reviews in
  Control}, 2020.

\bibitem{chisholm2018implications}
R.~H. Chisholm, P.~T. Campbell, Y.~Wu, S.~Y. Tong, J.~McVernon, and N.~Geard,
  ``Implications of asymptomatic carriers for infectious disease transmission
  and control,'' \emph{Royal Society open science}, vol.~5, no.~2, p. 172341,
  2018.

\bibitem{sandholm2010population}
W.~H. Sandholm, \emph{Population games and evolutionary dynamics}.\hskip 1em
  plus 0.5em minus 0.4em\relax MIT Press, 2010.

\bibitem{elokda2021dynamic}
\BIBentryALTinterwordspacing
E.~Elokda, A.~Censi, and S.~Bolognani, ``Dynamic population games,''
  \emph{arXiv preprint arXiv:2104.14662}, 2021. [Online]. Available:
  \url{https://arxiv.org/abs/2104.14662}
\BIBentrySTDinterwordspacing

\bibitem{jovanovic1988anonymous}
B.~Jovanovic and R.~W. Rosenthal, ``Anonymous sequential games,'' \emph{Journal
  of Mathematical Economics}, vol.~17, no.~1, pp. 77--87, 1988.

\bibitem{adlakha2015equilibria}
S.~Adlakha, R.~Johari, and G.~Y. Weintraub, ``Equilibria of dynamic games with
  many players: Existence, approximation, and market structure,'' \emph{Journal
  of Economic Theory}, vol. 156, pp. 269--316, 2015.

\bibitem{huang2006large}
M.~Huang, R.~P. Malham{\'e}, P.~E. Caines \emph{et~al.}, ``Large population
  stochastic dynamic games: closed-loop mckean-vlasov systems and the nash
  certainty equivalence principle,'' \emph{Communications in Information \&
  Systems}, vol.~6, no.~3, pp. 221--252, 2006.

\bibitem{gomes2010discrete}
D.~A. Gomes, J.~Mohr, and R.~R. Souza, ``Discrete time, finite state space mean
  field games,'' \emph{Journal de math{\'e}matiques pures et appliqu{\'e}es},
  vol.~93, no.~3, pp. 308--328, 2010.

\bibitem{neumann2020stationary}
B.~A. Neumann, ``Stationary equilibria of mean field games with finite state
  and action space,'' \emph{Dynamic Games and Applications}, 2020.

\bibitem{newman2010networks}
M.~Newman, \emph{Networks: {A}n {I}ntroduction}.\hskip 1em plus 0.5em minus
  0.4em\relax Oxford University Press, 2010.

\bibitem{filar2012competitive}
J.~Filar and K.~Vrieze, \emph{Competitive Markov decision processes}.\hskip 1em
  plus 0.5em minus 0.4em\relax Springer Science \& Business Media, 2012.

\bibitem{elokdadynamic_cdc_arxiv}
E.~Elokda, S.~Bolognani, and A.~R. Hota, ``A dynamic population model of
  strategic interaction and migration under epidemic risk,''
  \emph{PLACEHOLDER}, 2020.

\end{thebibliography}

\end{document}